\documentclass[conference]{IEEEtran}
\IEEEoverridecommandlockouts
\makeatletter
\def\conferenceheadnote{%
  \parbox[t]{\textwidth}{\centering\footnotesize
    Presented as a Tutorial at the 2026 IEEE International Conference on Blockchain and Cryptocurrency (ICBC), June 1, Brisbane, Australia\par}}
\def\ps@IEEEtitlepagestyle{%
  \def\@oddhead{\conferenceheadnote}
  \def\@evenhead{}
  \def\@oddfoot{}
  \def\@evenfoot{}}
\makeatother
\usepackage{tikz}
\usepackage{booktabs}
\usetikzlibrary{shadows,shapes.geometric}
\usepackage{xcolor}
\usepackage{array}

\usepackage[most]{tcolorbox}
\usepackage{colortbl}
\usepackage{makecell}

\definecolor{dopaCoral}{HTML}{FF6B6B}
\definecolor{dopaAmber}{HTML}{FF9F43}
\definecolor{dopaMint}{HTML}{10AC84}
\definecolor{dopaSky}{HTML}{54A0FF}
\definecolor{dopaLavender}{HTML}{5F27CD}
\definecolor{dopaPeach}{HTML}{FECA57}
\definecolor{dopaRose}{HTML}{FF9FF3}
\definecolor{dopaDark}{HTML}{2D3436}

\usepackage{framed}
\usepackage{wrapfig}
\definecolor{biobg}{HTML}{F2FAF5} 

\newenvironment{bioblock}
{%
 \MakeFramed{\advance\hsize-\width \FrameRestore}\noindent\footnotesize}
{\endMakeFramed}

\usepackage{cite}
\usepackage{amsmath,amssymb,amsfonts}
\usepackage{algorithmic}
\usepackage{graphicx}
\usepackage{textcomp}
\usepackage{url}
\usepackage{xurl}
\usepackage{enumitem}
\usepackage{booktabs}

\usepackage{array}

\definecolor{modblue}{RGB}{222,235,247}
\definecolor{modgreen}{RGB}{226,240,217}
\definecolor{modyellow}{RGB}{255,242,204}
\definecolor{modorange}{RGB}{252,228,214}
\definecolor{modpurple}{RGB}{234,209,220}
\definecolor{modgray}{RGB}{242,242,242}

\usepackage[table]{xcolor}
\usepackage{array}

\definecolor{cpgray}{HTML}{E9ECEF}

\definecolor{cpgreen}{HTML}{D4F1EC}

\definecolor{cppurpleA}{HTML}{E2D6F3} 
\definecolor{cppurpleB}{HTML}{E8DFF5} 
\definecolor{cppurpleC}{HTML}{EFE7FA} 

\definecolor{cpyellow}{HTML}{FFFCF0}

\definecolor{cphead}{RGB}{240,240,240}      
\definecolor{cpblue}{RGB}{226,236,244}      
\definecolor{cpgreen}{RGB}{228,240,232}     
\definecolor{cpsand}{RGB}{246,242,224}      
\definecolor{Cyellow}{HTML}{FFF6CC} 

\setlist[itemize]{leftmargin=*,nosep}

\title{Blockchain Infrastructure for Intelligent Cyber--Physical--Social Systems:\\
Post-Quantum Security, Interoperability, and Trustworthy Data Economies in the Era of Embodied AI}

\author{
\resizebox{\textwidth}{!}{
\begin{tabular}{ccccc}

\begin{tabular}{c}
Song Guo\textsuperscript{$\dagger$}\\
Hong Kong University of\\
Science and Technology\\
Hong Kong, China\\
songguo@cse.ust.hk
\end{tabular}
&
\begin{tabular}{c}
Huawei Huang\textsuperscript{$\dagger$}\\
Sun Yat-sen University\\
Zhuhai, China\\
huanghw28@mail.sysu.edu.cn
\end{tabular}
&
\begin{tabular}{c}
Dongping Liu\textsuperscript{$\dagger$}\\
Amazon Web Services\\
Hong Kong, China\\
dpliu@amazon.com
\end{tabular}
&
\begin{tabular}{c}
Aoyu Zhang\textsuperscript{$\dagger$}\\
Amazon Web Services\\
Beijing, China\\
aoyuzhan@amazon.com
\end{tabular}
&
\begin{tabular}{c}
Luyao Zhang\textsuperscript{$\dagger$}\\
Duke Kunshan University\\
Suzhou, China\\
lz183@duke.edu
\end{tabular}

\end{tabular}
}
\thanks{\textsuperscript{$\dagger$}The authors are listed in alphabetical order by last name.}
}

\begin{document}

\maketitle

\begin{figure*}[!htbp]
\centering
\includegraphics[width=\textwidth]{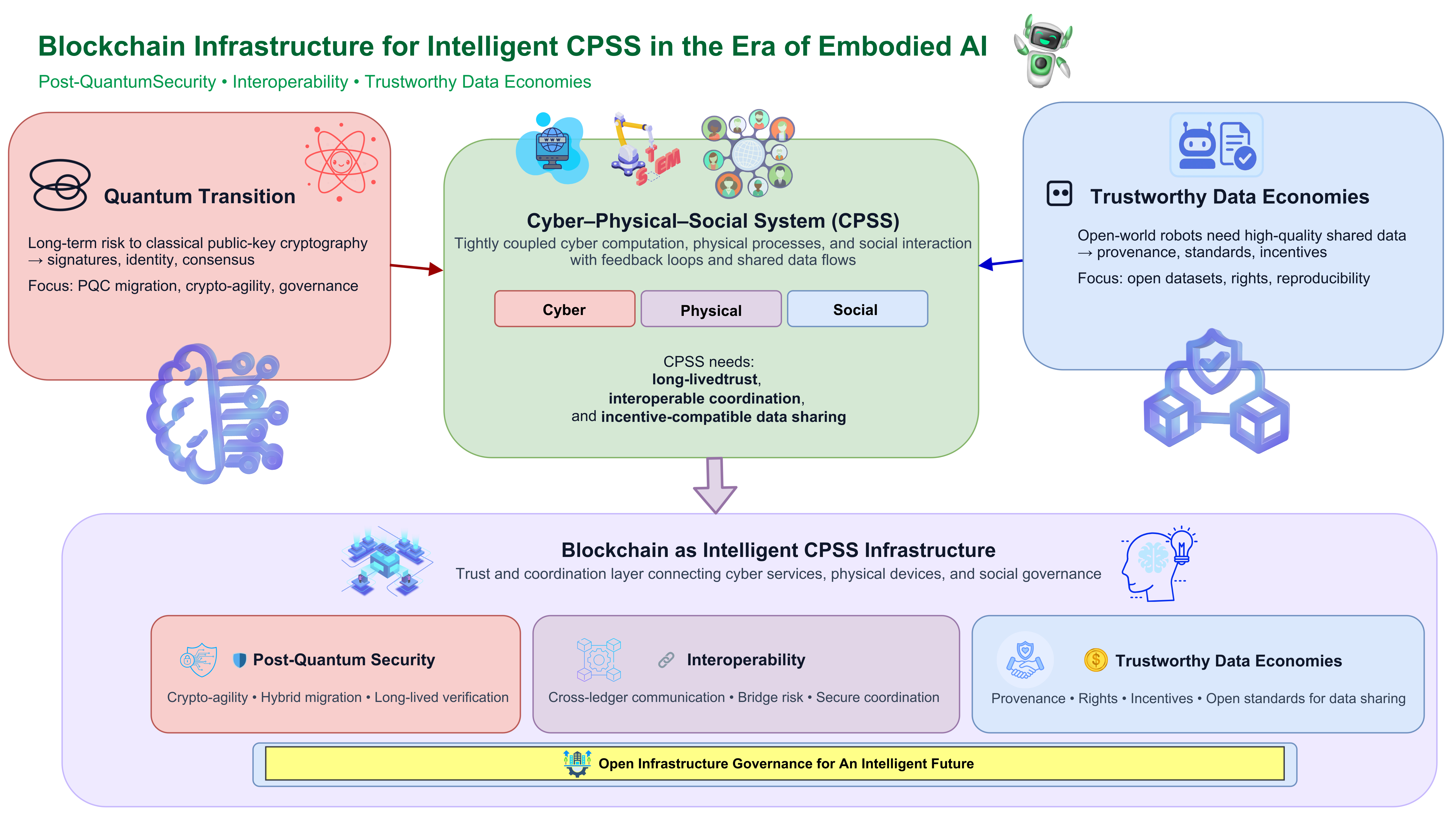}
\caption{Overview of the Quantum-Blockchain-Embodied (QBE) infrastructure paradigm illustrating the relationship between quantum transition, embodied AI data economies, and cyber--physical--social systems (CPSS).}
\label{fig:qbe_teaser}
\end{figure*}


\begin{abstract}
The deployment of embodied artificial intelligence via world-model-based robotic systems presents a transformative opportunity for blockchain infrastructure, establishing urgent demand for trustworthy data provenance, cross-organizational governance, and incentive-compatible sharing across decentralized ecosystems. Simultaneously, advances in quantum computing—recognized by the 2025 Nobel Prize in Physics and the Turing Award—threaten the cryptographic primitives securing these emerging data economies, creating an interdependent infrastructure imperative: the long-lived verification required by embodied AI systems depends entirely on crypto-agile architectures capable of withstanding quantum adversaries.
This tutorial examines blockchain as the essential coordination layer bridging this dual transition, evolving from financial substrate to foundational Cyber-Physical-Social Systems infrastructure that simultaneously secures against quantum cryptanalysis and enables scalable, trustworthy data economies. Unlike conventional surveys, the session opens with an immersive AWS Braket demonstration that engages participants with superconducting, trapped-ion, and neutral-atom hardware to empirically assess cryptographic threat timelines and witness signature transitions from ECDSA to post-quantum schemes. This foundation supports five integrated modules progressing from embodied AI and world-model requirements through quantum hardware reality and evidence-based security migration, to scalable cross-shard architectures via BrokerChain protocols, trustworthy data economies implementing Croissant metadata standards and robotic learning provenance, and industry ecosystem integration for multi-modal cloud deployment. Each module transitions logically to form a cohesive pipeline from empirical security assessment through scalable coordination to production-ready governance. By bridging quantum hardware realities with embodied AI data requirements, this tutorial charts the evolution of blockchain as the unified infrastructure for next-generation decentralized intelligent environments, providing participants with open-source experimental frameworks and actionable roadmaps for architecting quantum-resistant, interoperable, and data-trustworthy systems.
\end{abstract}


\section{Objectives and Motivation}

IEEE ICBC emphasizes emerging research directions that reshape blockchain infrastructure. This tutorial addresses the convergence of quantum computing and embodied artificial intelligence recently accelerated by two scientific milestones: quantum computing advances recognized by the 2025 Nobel Prize in Physics and the Turing Award \cite{nobel2025quantum,turing2026quantum}, and the deployment of embodied AI systems operating via world models in open-world environments \cite{monwilliams2025embodied,Zhu_2025_ICCV}.
As illustrated in Figure~\ref{fig:qbe_teaser}, these transitions create convergent demands on Cyber--Physical--Social Systems (CPSS). Quantum computing introduces long-term risks to classical public-key cryptography---threatening signatures, identity, and consensus mechanisms \cite{fedorov2018quantum,google2025otoc}---while embodied AI generates massive interaction data requiring trustworthy sharing, provenance, and incentive-compatible governance \cite{monwilliams2025embodied}.
Blockchain must therefore evolve as the trust and coordination layer connecting cyber services, physical devices, and social governance, simultaneously providing: (1) Post-Quantum Security via crypto-agility and hybrid migration; (2) Interoperability for cross-ledger coordination; and (3) Trustworthy Data Economies enabling open datasets with rights management, interoperability, and reproducibility. This tutorial provides a unified perspective on architecting this infrastructure for long-lived verification and decentralized intelligent environments. All tutorial materials---including slides, code notebooks, and demonstration scripts---are openly accessible on GitHub.\footnote{\url{https://github.com/QuantBlockchain/ieee-icbc-tutorial-qbe}}

\section{Timeline and Intended Audience}

\textbf{Duration:} 2 hours, comprising an opening interactive hands-on demonstration using AWS Braket, five sequential technical modules (M1--M5), and a closing synthesis panel.

\textit{Intended Audience:} This tutorial targets an interdisciplinary audience including researchers in blockchain, security and cryptography, distributed systems, artificial intelligence, and token economy, as well as industry practitioners. It provides a unified perspective on how quantum-resistant security, interoperable blockchain architectures, and trustworthy data economies jointly shape emerging Intelligent CPSS, progressing from quantum-hardware exploration through embodied AI infrastructure to industry ecosystem deployment.

\section{Timeliness and Technical Issues}

The simultaneous transition toward quantum-capable computation and large-scale embodied AI deployment creates a critical juncture where infrastructure decisions will determine whether next-generation ecosystems can resist quantum cryptanalysis while sustaining data-driven AI economies. This tutorial addresses three integrated technical pillars (Figure~\ref{fig:qbe_teaser}):

\begin{itemize}
\item \textbf{Post-Quantum Security}: Empirical threat assessment via Amazon Braket (coherence times, gate fidelities, error rates) and crypto-agility mechanisms enabling hybrid migration and long-lived verification \cite{fedorov2018quantum,nobel2025quantum,turing2026quantum,google2025otoc};

\item \textbf{Interoperability}: BrokerChain cross-shard protocols and bridge security for heterogeneous cross-ledger communication required for CPSS-scale deployment \cite{10.14778/3587136.3587143,huang2022brokerchain,huang2025brokerchain,augusto2024sok};

\item \textbf{Trustworthy Data Economies}: Croissant metadata standards, blockchain provenance for robotic datasets (LET, Kuavo), and incentive-compatible mechanisms ensuring high-quality data sharing with open standards and reproducibility \cite{NEURIPS2024_9547b09b,yan2025ethbeacon,chemaya2025uniswap,10704461,o2024open,letdataset,kuavochallenge}.
\end{itemize}

Open challenges include governance for cryptographic transitions, standards-aware interoperability, and incentive alignment, sustaining long-term collaboration in Intelligent CPSS.

\section{Tutorial Speakers and Biographies}
\label{sec:contributors}

The speakers bring extensive interdisciplinary experience across blockchain systems, machine learning infrastructure, quantum computing, and economic mechanism design, building on prior individual tutorial and workshop experience—including a collaborative tutorial accepted at WWW 2026 on quantum-safe and AI-enhanced blockchains for the Web—while extending this work toward a new integrated perspective on post-quantum security, interoperable blockchain infrastructure, and trustworthy data economies for embodied AI in Intelligent CPSS.

\noindent
\footnotetext[2]{\url{https://cse.hkust.edu.hk/~songguo/}}
\begin{bioblock}

\begin{wrapfigure}{l}{0.25\textwidth}
\includegraphics[width=\linewidth]{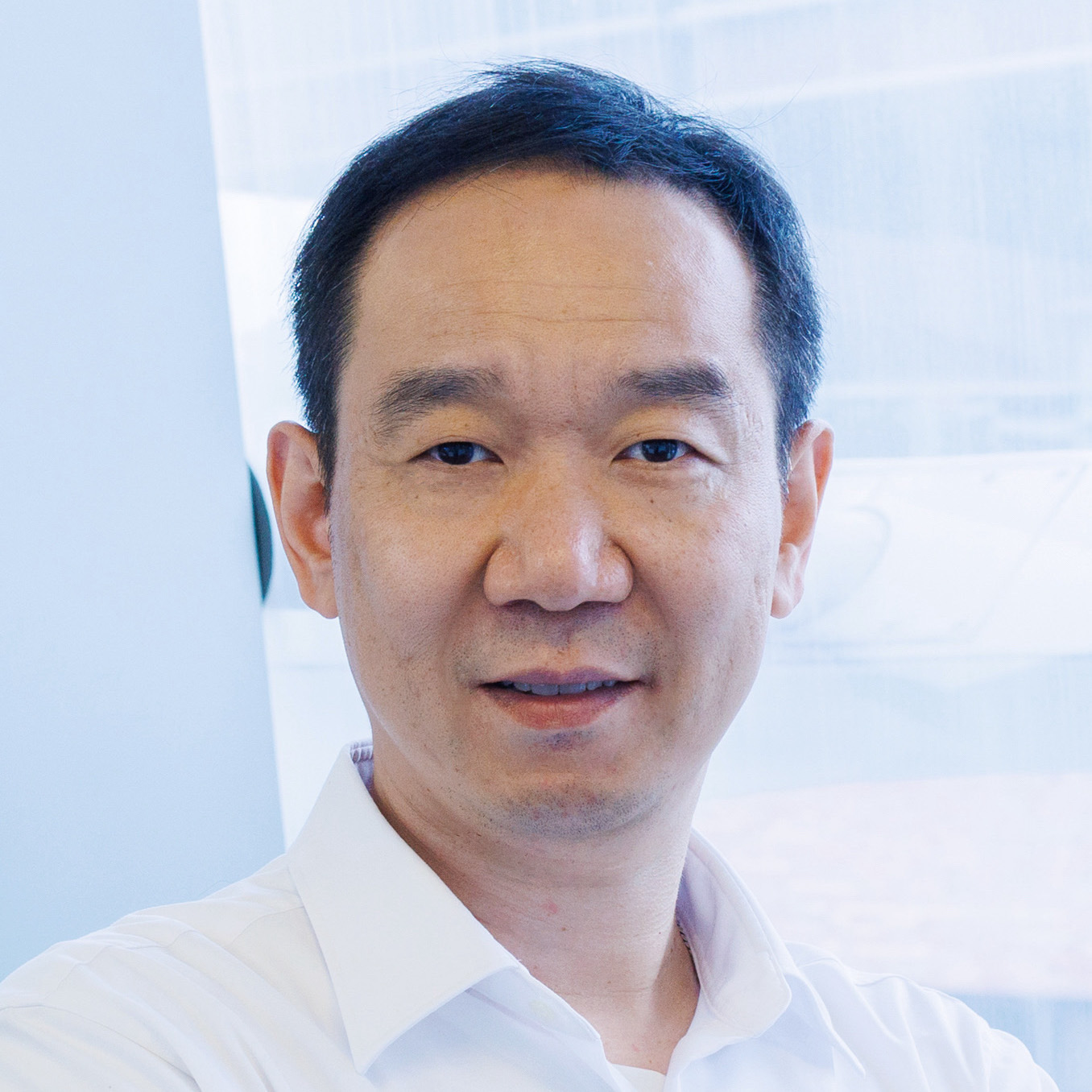}
\end{wrapfigure}

\textbf{Song Guo}\footnotemark[2] is a Chair Professor at HKUST. He also holds a Changjiang Chair Professorship awarded by the Ministry of Education of China. His research focuses on Large Language Models, Edge AI, and Machine Learning Systems \cite{10.1145/3627703.3629567,10.14778/3587136.3587143,Zhu_2025_ICCV}. A Highly Cited Researcher, he has received over a dozen Best Paper Awards and the Edward J. McCluskey Technical Achievement Award (2024), First Prize in Natural Science (China Electronics Society, 2023), and Gold Medals at Geneva Inventions Expo (2023, 2024). He is a Fellow of the Canadian Academy of Engineering, Member of Academia Europaea, Fellow of the IEEE, Distinguished Member of the ACM, and Fellow of the Asia-Pacific Artificial Intelligence Association. He served as IEEE Communications Society Distinguished Lecturer and Board of Governors member. He is Editor-in-Chief of IEEE Transactions on Cloud Computing and founding Editor-in-Chief of IEEE Open Journal of the Computer Society. He has chaired numerous IEEE/ACM conference committees, delivered 100+ keynotes, and serves as Secretary General of CCF Hong Kong.

\medskip

\begin{wrapfigure}{l}{0.25\textwidth}
\includegraphics[width=\linewidth]{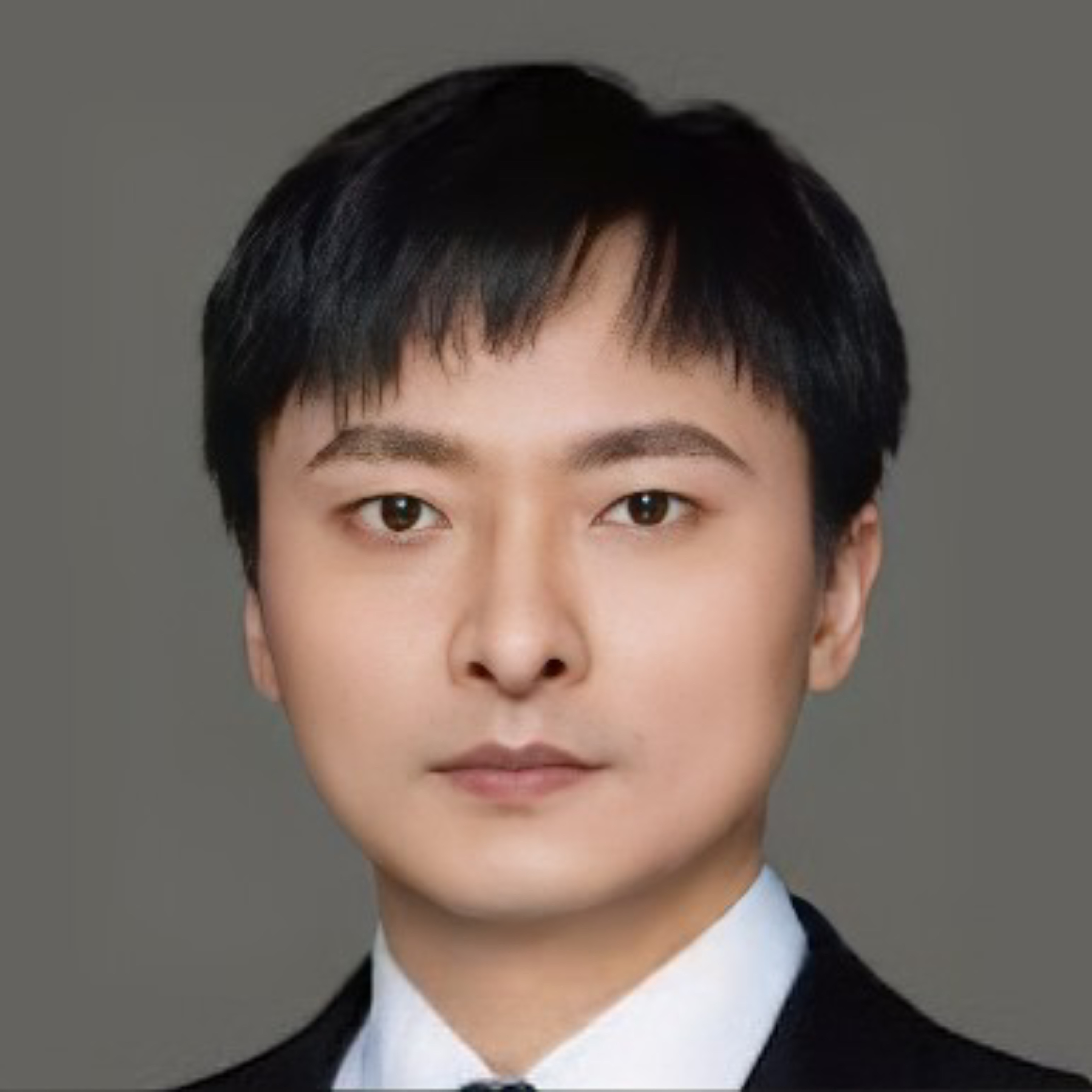}
\end{wrapfigure}

\textbf{Huawei Huang}\footnotemark[3] received the Ph.D. degree from the University of Aizu, Japan, in 2016. He is a Professor at Sun Yat-sen University, with dual appointments at Lingnan College and the Hong Kong Institute of Advanced Studies. He previously served as a JSPS Research Fellow and Assistant Professor at Kyoto University. He is a recipient of the Guangdong Outstanding Young Scholar Fund, an IEEE Senior Member, and a CCF Senior Member. Since 2021, he has been listed in the Stanford University World's Top 2\% Scientists ranking. His research interests include high-performance blockchain systems, DeFi protocols, and Web3 infrastructures. He has published in top-tier venues such as IEEE ToN, JSAC, TPDS, INFOCOM, and WWW, with over 9,000 citations. He has served as chair for more than ten international conferences and led over ten national research projects, including the National Key R\&D Program of China and NSFC programs. He has authored three academic books on blockchain, one textbook, and one popular science book. He leads the development of BlockEmulator, an open-source blockchain experimental platform supporting researchers in over 90 countries, and the BrokerChain\cite{huang2025brokerchain} project, which launched its Academic Testnet in June 2025.

\medskip

\begin{wrapfigure}{l}{0.25\textwidth}
\includegraphics[width=\linewidth]{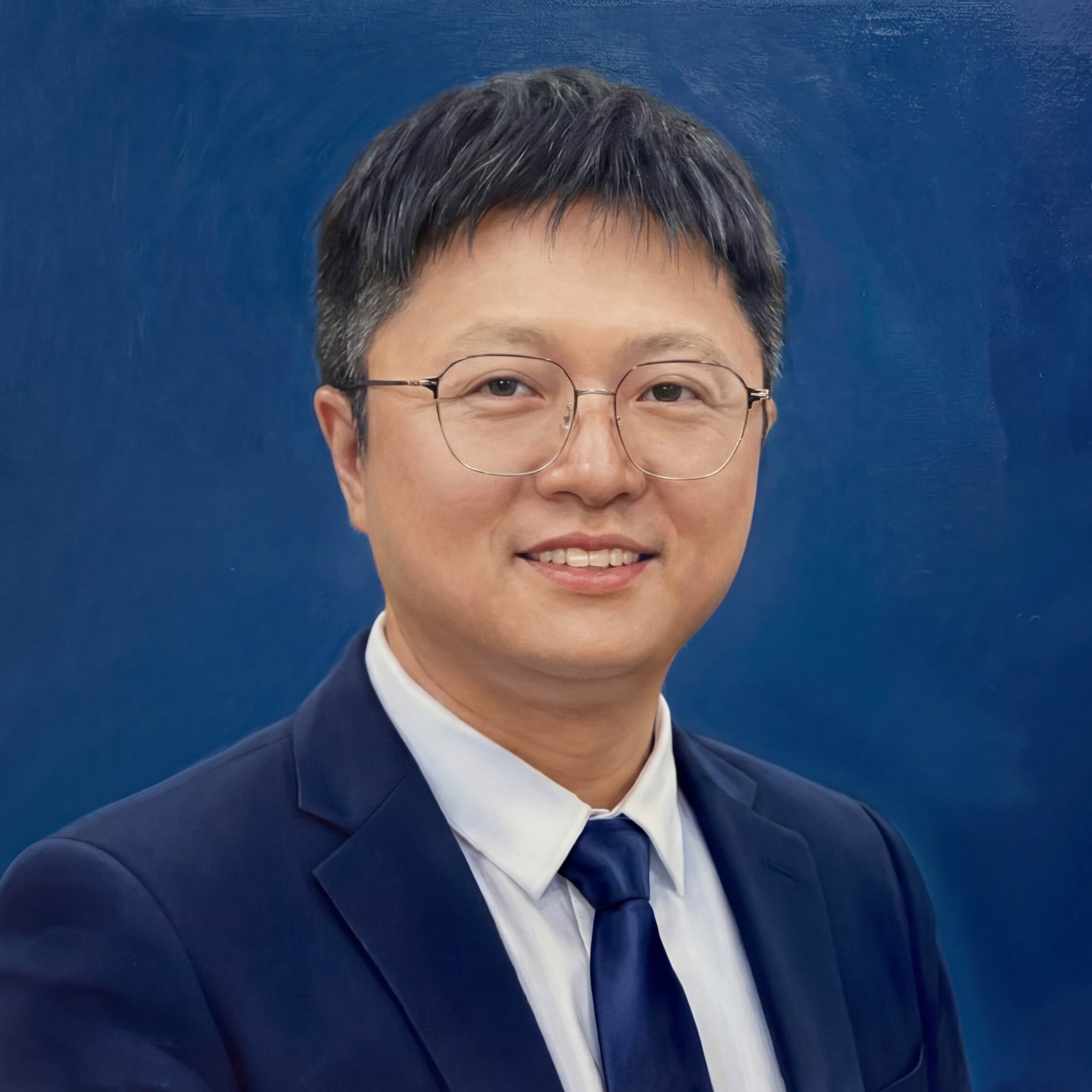}
\end{wrapfigure}
\textbf{Dongping Liu}\footnotemark[4] is Senior Industry Business Development Manager for Higher Education and Research at Amazon Web Services,where he specializes in driving digital transformation in academia through AI, deep learning, and high-performance computing solutions. With 15 years of experience in the higher education and research sector, he has successfully led the implementation of cutting-edge cloud technologies for educational institutions and research organizations. His expertise spans educational technology innovation, business development strategy, and large-scale research computing deployments. He has been awarded the First Prize of Beijing Science and Technology Award, published 30+ research papers, and obtained 10+ authorized patents along with multiple software copyrights. He contributes to this tutorial by bridging academic research needs with practical AWS cloud solutions and sharing insights on AI/HPC applications in higher education. He holds a Ph.D. in Physics from the Institute of Physics, Chinese Academy of Sciences, and completed postdoctoral research at McGill University, Canada.

\medskip

\begin{wrapfigure}{l}{0.25\textwidth}
\includegraphics[width=\linewidth]{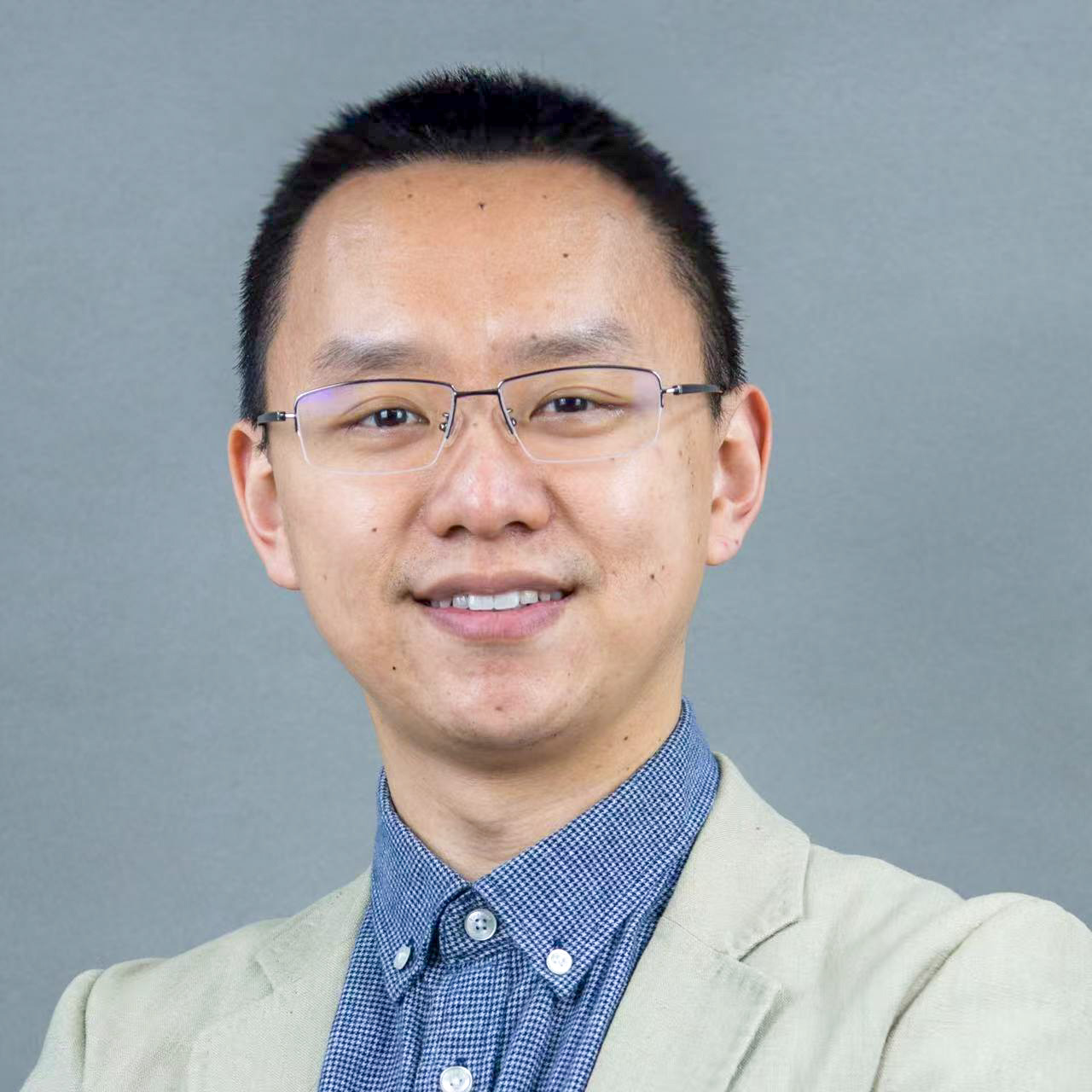}
\end{wrapfigure}
\textbf{Aoyu Zhang}\footnotemark[5] is Senior Applied Scientist at AWS China Solution Development Center, where he advances the design and implementation of AI and quantum cloud solutions. He used to develop open source quantum computing solutions for drug discovery under Amazon Braket. His research also includes hybrid quantum–classical algorithms, quantum benchmarking, and cross-domain applications in blockchain security and Web-scale cryptography. He delivered the keynote at the 2025 forum on Cloud Quantum Computing Services for Blockchain Security and contributes to this tutorial by leading the quantum core and live Amazon Braket demonstrations. He holds a Ph.D. in Biomedical Engineering in Peking University.

\medskip

\begin{wrapfigure}{l}{0.25\textwidth}
\includegraphics[width=\linewidth]{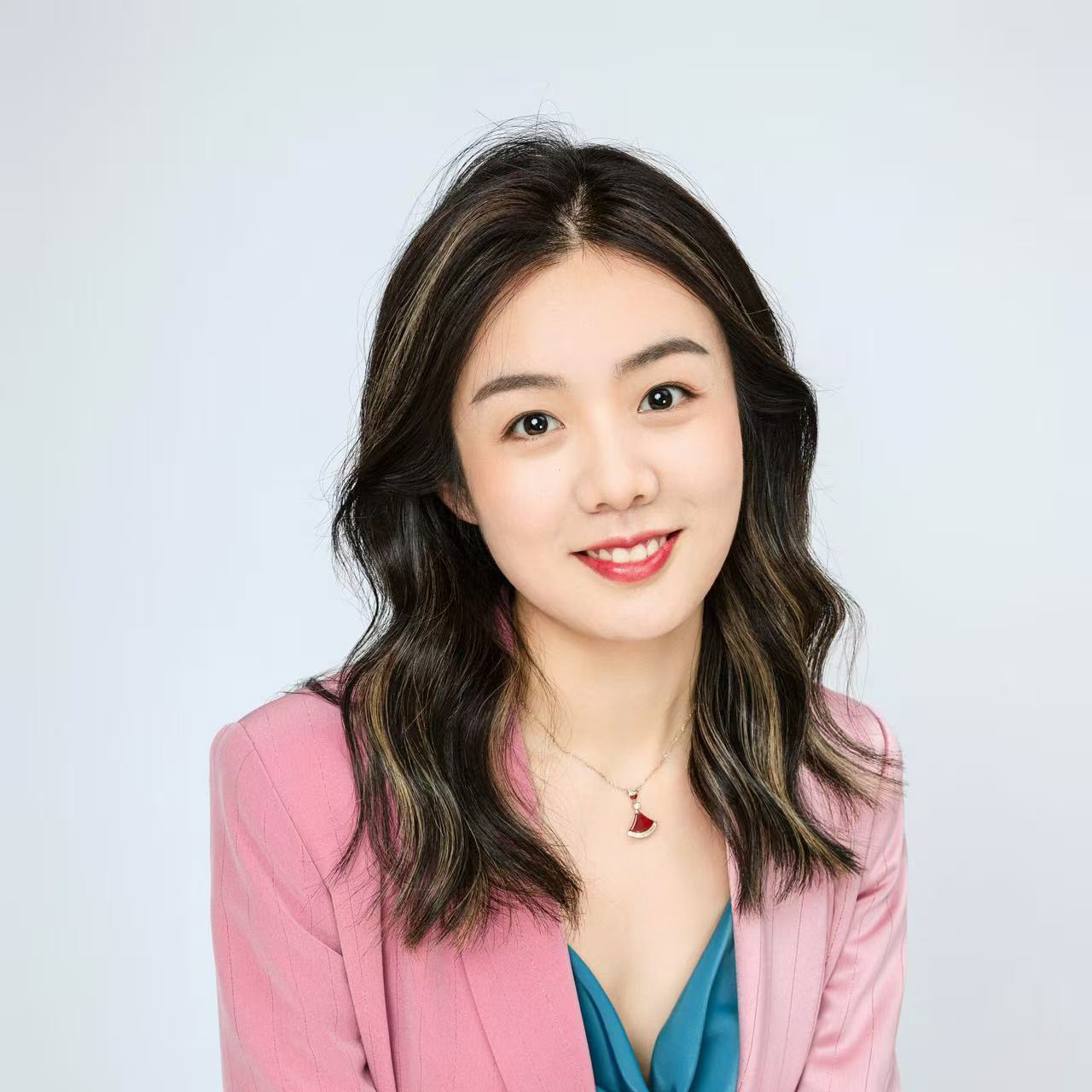}
\end{wrapfigure}

\textbf{Luyao Zhang}\footnotemark[6] is a tenure-track Assistant Professor of Economics and Senior Research Scientist at Duke Kunshan University. Her research bridges computational and economic sciences through groundbreaking technologies for intelligent economics, encompassing big data, blockchain, generative AI, and geospatial analysis, with publications in \emph{Review of Economics and Statistics}\cite{levin2022bridging}, \emph{Scientific Data}\cite{yan2025ethbeacon,chemaya2025uniswap,liu2022bitcoincohort}, \emph{ACM CCS}\cite{liu2022eip1559}, \emph{IEEE S\&P}\cite{augusto2024sok}, \emph{ACM CSCW}\cite{xiao2024nft}, and \emph{NeurIPS Datasets and Benchmarks}\cite{NEURIPS2024_9547b09b}. She holds a Ph.D.\ from The Ohio State University (Presidential Fellowship, NSF support) and a B.A./B.S.\ dual degree from Peking University, with professional certificates from Oxford and MIT in blockchain, reinforcement learning, and quantum computing. Her NSFC-funded project ``Trust Mechanism Design on Blockchain'' integrates game theory, reinforcement learning, and human--AI interaction. She serves as Guest Editor for \emph{Electronic Markets}\footnote{https://link.springer.com/collections/hjiffhehga}, Editorial Board Member of \emph{Scientific Data}\footnote{https://www.nature.com/sdata/editorial-board} and \emph{Blockchain: Research and Applications}\footnote{https://www.sciencedirect.com/journal/blockchain-research-and-applications/about/editorial-board}, Working Group Secretary for IEEE P3469, and Academic Editor of \emph{Blockchain --- Pioneering the Web3 Infrastructure for an Intelligent Future} (2025)\cite{Zhang_2025}.

\end{bioblock}

\footnotetext[3]{\url{http://xintelligence.pro/}}
\footnotetext[4]{\url{https://www.linkedin.com/in/dpliu/}}
\footnotetext[5]{\url{https://www.linkedin.com/in/aoyu-zhang-quantum/}}
\footnotetext[6]{\url{https://scholars.duke.edu/person/luyao.zhang}}

\section{Outline of Tutorial Content}

The tutorial is structured as an opening interactive demonstration, five sequential technical modules, and a closing synthesis panel (Table~\ref{tab:tutorial_outline}). 

\vspace{4pt}

\newcommand{\iconQuantum}[1]{%
\begin{tikzpicture}[baseline=-3pt, scale=0.55]
    \draw[#1, line width=1pt] (0,0) circle (0.35);
    \draw[#1, line width=0.8pt] (0,0) ellipse (0.55 and 0.2);
    \draw[#1, line width=0.8pt, rotate=60] (0,0) ellipse (0.55 and 0.2);
    \draw[#1, line width=0.8pt, rotate=-60] (0,0) ellipse (0.55 and 0.2);
    \fill[#1] (0,0) circle (0.1);
\end{tikzpicture}%
}

\newcommand{\iconRobot}[1]{%
\begin{tikzpicture}[baseline=-3pt, scale=0.55]
    \draw[#1, line width=1pt, rounded corners=1.5pt] (-0.3,-0.15) rectangle (0.3,0.25);
    \draw[#1, line width=0.8pt, rounded corners=1pt] (-0.15,0.25) -- (-0.15,0.4) -- (0.15,0.4) -- (0.15,0.25);
    \fill[#1] (-0.12,0.08) circle (0.06);
    \fill[#1] (0.12,0.08) circle (0.06);
    \draw[#1, line width=0.8pt] (-0.18,-0.05) -- (0.18,-0.05);
    \draw[#1, line width=0.8pt, rounded corners=1pt] (-0.35,-0.15) -- (-0.35,0.05) -- (-0.3,0.05);
    \draw[#1, line width=0.8pt, rounded corners=1pt] (0.35,-0.15) -- (0.35,0.05) -- (0.3,0.05);
\end{tikzpicture}%
}

\newcommand{\iconShield}[1]{%
\begin{tikzpicture}[baseline=-3pt, scale=0.55]
    \draw[#1, line width=1pt, rounded corners=1pt] 
        (-0.3,-0.1) -- (-0.3,0.15) -- (0,0.4) -- (0.3,0.15) -- (0.3,-0.1) -- (0,-0.25) -- cycle;
    \draw[#1, line width=0.8pt] (-0.12,0.05) -- (0,0.18) -- (0.18,-0.05);
\end{tikzpicture}%
}

\newcommand{\iconNetwork}[1]{%
\begin{tikzpicture}[baseline=-3pt, scale=0.55]
    \fill[#1] (0,0.25) circle (0.1);
    \fill[#1] (-0.25,-0.15) circle (0.1);
    \fill[#1] (0.25,-0.15) circle (0.1);
    \draw[#1, line width=0.8pt] (0,0.25) -- (-0.25,-0.15);
    \draw[#1, line width=0.8pt] (0,0.25) -- (0.25,-0.15);
    \draw[#1, line width=0.8pt] (-0.25,-0.15) -- (0.25,-0.15);
\end{tikzpicture}%
}

\newcommand{\iconData}[1]{%
\begin{tikzpicture}[baseline=-3pt, scale=0.55]
    \draw[#1, line width=0.8pt] (-0.3,-0.2) rectangle (0.3,0.3);
    \draw[#1, line width=1pt, fill=#1!20] (-0.2,-0.2) rectangle (-0.05,0.1);
    \draw[#1, line width=1pt, fill=#1!20] (0.0,-0.2) rectangle (0.15,-0.05);
    \draw[#1, line width=1pt, fill=#1!20] (0.18,-0.2) rectangle (0.28,0.2);
    \draw[#1, line width=0.6pt, dashed] (-0.25,0.22) -- (0.25,0.22);
\end{tikzpicture}%
}

\newcommand{\iconCloud}[1]{%
\begin{tikzpicture}[baseline=-3pt, scale=0.55]
    \draw[#1, line width=0.9pt] (-0.2,0.15) arc (180:0:0.2);
    \draw[#1, line width=0.9pt] (-0.35,0.05) arc (180:0:0.15);
    \draw[#1, line width=0.9pt] (0.05,0.05) arc (180:0:0.15);
    \draw[#1, line width=0.9pt] (-0.35,0.05) -- (-0.35,-0.05);
    \draw[#1, line width=0.9pt] (0.35,0.05) -- (0.35,-0.05);
    \draw[#1, line width=0.9pt] (-0.2,-0.15) arc (180:360:0.2);
    \fill[#1] (0,-0.05) circle (0.1);
    \draw[white, line width=0.6pt] (0,-0.05) circle (0.05);
\end{tikzpicture}%
}

\newcommand{\iconStar}[1]{%
\begin{tikzpicture}[baseline=-3pt, scale=0.55]
    \node[star, star points=5, star point ratio=2, draw=#1, fill=#1!20, 
          line width=0.9pt, minimum size=0.7cm, inner sep=0pt] at (0,0) {};
\end{tikzpicture}%
}


\begin{table*}[htbp]
\centering
\caption{Tutorial Outline and Presenter Assignment}
\label{tab:tutorial_outline}
\small
\setlength{\tabcolsep}{0pt}

\begin{tcolorbox}[
    enhanced,
    colback=dopaDark!90,
    colframe=dopaDark,
    boxrule=1pt,
    arc=2mm,
    left=7pt, right=7pt, top=5pt, bottom=5pt,
    fuzzy shadow={1mm}{-0.6mm}{0mm}{0.3mm}{dopaDark!60},
    sidebyside,
    sidebyside align=top,
    lefthand width=2.6cm,
    segmentation style={draw=white!40, line width=0.5pt},
    fontupper=\bfseries\small\color{white},
    fontlower=\bfseries\small\color{white},
]
Module\par\vspace{4pt}
{\footnotesize Presenter}
\tcblower
Topic\par\vspace{4pt}
{\footnotesize Content and Progression}
\end{tcolorbox}

\vspace{4pt}

\begin{tcolorbox}[
    enhanced,
    colback=dopaCoral!6,
    colframe=dopaCoral!70!black,
    boxrule=0.8pt,
    arc=2.5mm,
    left=7pt, right=7pt, top=5pt, bottom=5pt,
    fuzzy shadow={1mm}{-0.6mm}{0mm}{0.3mm}{dopaCoral!30!black},
    sidebyside,
    sidebyside align=top,
    lefthand width=2.6cm,
    segmentation style={draw=dopaCoral!45, line width=0.5pt, dash pattern={on 3pt off 2pt}},
    fontupper=\small, fontlower=\small,
    collower=black!85,
    before upper={
        \centering\iconQuantum{dopaCoral}\par\vspace{2pt}
        \textbf{\color{dopaCoral!85!black}Opening Demo $\triangleright$}\par\vspace{1pt}
        {\footnotesize\color{black!65}Dongping Liu,\\Aoyu Zhang,\\Luyao Zhang}\par\vspace{2pt}
    },
    before lower={
        \textbf{\color{dopaCoral!80!black}Quantum-Web3-AI Convergence}\par\vspace{2pt}
    }
]
\tcblower
Hands-on AWS Braket demonstration of quantum computing advances recognized by the 2025 Nobel Prize and Turing Award, empirical threat assessment to blockchain ECDSA signatures, and post-quantum signature transition pathways, \textit{establishing the empirical CPSS framework that motivates the technical modules}.
\end{tcolorbox}

\vspace{3pt}

\begin{tcolorbox}[
    enhanced,
    colback=dopaAmber!6,
    colframe=dopaAmber!70!black,
    boxrule=0.8pt,
    arc=2.5mm,
    left=7pt, right=7pt, top=5pt, bottom=5pt,
    fuzzy shadow={1mm}{-0.6mm}{0mm}{0.3mm}{dopaAmber!30!black},
    sidebyside,
    sidebyside align=top,
    lefthand width=2.6cm,
    segmentation style={draw=dopaAmber!45, line width=0.5pt, dash pattern={on 3pt off 2pt}},
    fontupper=\small, fontlower=\small,
    collower=black!85,
    before upper={
        \centering\iconRobot{dopaAmber}\par\vspace{2pt}
        \textbf{\color{dopaAmber!90!black}M1 $\triangleright$}\par\vspace{1pt}
        {\footnotesize\color{black!65}Song Guo}\par\vspace{2pt}
    },
    before lower={
        \textbf{\color{dopaAmber!85!black}Embodied AI and World Models}\par\vspace{2pt}
    }
]
\tcblower
Presentation of edge-distributed robotic systems, world-model-based architectures (IRASim) and world model-based policy optimization for vision-language-action models (WMPO), \textit{defining the target infrastructure that M2 must secure against quantum threats}.
\end{tcolorbox}

\vspace{3pt}

\begin{tcolorbox}[
    enhanced,
    colback=dopaMint!5,
    colframe=dopaMint!80!black,
    boxrule=0.8pt,
    arc=2.5mm,
    left=7pt, right=7pt, top=5pt, bottom=5pt,
    fuzzy shadow={1mm}{-0.6mm}{0mm}{0.3mm}{dopaMint!35!black},
    sidebyside,
    sidebyside align=top,
    lefthand width=2.6cm,
    segmentation style={draw=dopaMint!50, line width=0.5pt, dash pattern={on 3pt off 2pt}},
    fontupper=\small, fontlower=\small,
    collower=black!85,
    before upper={
        \centering\iconShield{dopaMint}\par\vspace{2pt}
        \textbf{\color{dopaMint!90!black}M2 $\triangleright$}\par\vspace{1pt}
        {\footnotesize\color{black!65}Aoyu Zhang}\par\vspace{2pt}
    },
    before lower={
        \textbf{\color{dopaMint!85!black}Quantum Hardware Threat Assessment}\par\vspace{2pt}
    }
]
\tcblower
Empirical analysis of quantum capabilities via AWS Braket and evidence-based post-quantum migration strategies, \textit{determining the security requirements that M3 must address through scalable architectures}.
\end{tcolorbox}

\vspace{3pt}

\begin{tcolorbox}[
    enhanced,
    colback=dopaSky!6,
    colframe=dopaSky!75!black,
    boxrule=0.8pt,
    arc=2.5mm,
    left=7pt, right=7pt, top=5pt, bottom=5pt,
    fuzzy shadow={1mm}{-0.6mm}{0mm}{0.3mm}{dopaSky!35!black},
    sidebyside,
    sidebyside align=top,
    lefthand width=2.6cm,
    segmentation style={draw=dopaSky!45, line width=0.5pt, dash pattern={on 3pt off 2pt}},
    fontupper=\small, fontlower=\small,
    collower=black!85,
    before upper={
        \centering\iconNetwork{dopaSky}\par\vspace{2pt}
        \textbf{\color{dopaSky!90!black}M3 $\triangleright$}\par\vspace{1pt}
        {\footnotesize\color{black!65}Huawei Huang}\par\vspace{2pt}
    },
    before lower={
        \textbf{\color{dopaSky!85!black}Scalable Arch. and Interoperability}\par\vspace{2pt}
    }
]
\tcblower
Demonstration of BrokerChain cross-shard protocols and cross-ledger coordination, \textit{providing the scalable foundation upon which M4 implements trustworthy data economies}.
\end{tcolorbox}

\vspace{3pt}

\begin{tcolorbox}[
    enhanced,
    colback=dopaLavender!5,
    colframe=dopaLavender!80!black,
    boxrule=0.8pt,
    arc=2.5mm,
    left=7pt, right=7pt, top=5pt, bottom=5pt,
    fuzzy shadow={1mm}{-0.6mm}{0mm}{0.3mm}{dopaLavender!35!black},
    sidebyside,
    sidebyside align=top,
    lefthand width=2.6cm,
    segmentation style={draw=dopaLavender!50, line width=0.5pt, dash pattern={on 3pt off 2pt}},
    fontupper=\small, fontlower=\small,
    collower=black!85,
    before upper={
        \centering\iconData{dopaLavender}\par\vspace{2pt}
        \textbf{\color{dopaLavender}M4 $\triangleright$}\par\vspace{1pt}
        {\footnotesize\color{black!65}Luyao Zhang}\par\vspace{2pt}
    },
    before lower={
        \textbf{\color{dopaLavender!95}Trustworthy Data Economies}\par\vspace{2pt}
    }
]
\tcblower
Implementation of Croissant metadata standards for ML-ready datasets, security analysis of blockchain interoperability mechanisms, and blockchain provenance for Ethereum Beacon Chain and Uniswap data, \textit{establishing secure cross-ledger data exchange protocols that leverage M3's scalable foundation}.
\end{tcolorbox}

\vspace{3pt}

\begin{tcolorbox}[
    enhanced,
    colback=dopaPeach!8,
    colframe=dopaPeach!70!black,
    boxrule=0.8pt,
    arc=2.5mm,
    left=7pt, right=7pt, top=5pt, bottom=5pt,
    fuzzy shadow={1mm}{-0.6mm}{0mm}{0.3mm}{dopaPeach!30!black},
    sidebyside,
    sidebyside align=top,
    lefthand width=2.6cm,
    segmentation style={draw=dopaPeach!45, line width=0.5pt, dash pattern={on 3pt off 2pt}},
    fontupper=\small, fontlower=\small,
    collower=black!85,
    before upper={
        \centering\iconCloud{dopaPeach}\par\vspace{2pt}
        \textbf{\color{dopaPeach!90!black}M5 $\triangleright$}\par\vspace{1pt}
        {\footnotesize\color{black!65}Dongping Liu}\par\vspace{2pt}
    },
    before lower={
        \textbf{\color{dopaPeach!85!black}Industry Ecosystem Integration}\par\vspace{2pt}
    }
]
\tcblower
Operationalization of cloud infrastructure and multi-modal data deployment (e.g., LET, Kuavo) for embodied AI, \textit{completing the pipeline from theoretical security to production-ready CPSS infrastructure}.
\end{tcolorbox}

\vspace{3pt}

\begin{tcolorbox}[
    enhanced,
    colback=dopaRose!10,
    colframe=dopaRose!70!black,
    boxrule=0.8pt,
    arc=2.5mm,
    left=7pt, right=7pt, top=5pt, bottom=5pt,
    fuzzy shadow={1mm}{-0.6mm}{0mm}{0.3mm}{dopaRose!30!black},
    sidebyside,
    sidebyside align=top,
    lefthand width=2.6cm,
    segmentation style={draw=dopaRose!45, line width=0.5pt, dash pattern={on 3pt off 2pt}},
    fontupper=\small, fontlower=\small,
    collower=black!85,
    before upper={
        \centering\iconStar{dopaRose}\par\vspace{2pt}
        \textbf{\color{dopaRose!90!black}Panel Synthesis}\par\vspace{1pt}
        {\footnotesize\color{black!65}All Speakers}\par\vspace{2pt}
    },
    before lower={
        \textbf{\color{dopaRose!85!black}Integrated Roadmap}\par\vspace{2pt}
    }
]
\tcblower
Synthesis of quantum-resistant security, scalable architecture, and trustworthy data economies into a unified CPSS strategy and future research agenda.
\end{tcolorbox}

\end{table*}

The tutorial architecture follows a logical progression from empirical foundations to operational deployment. The \textbf{Opening Demo} establishes immediate hands-on familiarity with quantum computing advances recognized by the 2025 Nobel Prize in Physics and the Turing Award\cite{nobel2025quantum,turing2026quantum}, using AWS Braket\footnote{https://aws.amazon.com/braket/} to empirically demonstrate the concrete threat timeline to blockchain ECDSA signatures and witness transitions to post-quantum cryptographic schemes \cite{liu2026qsignaiquantumrandomnessseededidentitysignatures,liu2026quantumfuturesinteractivelive}. This immersive exploration of superconducting, trapped-ion, and neutral-atom modalities grounds participants in the empirical reality of quantum hardware capabilities and establishes the CPSS security framework that motivates subsequent technical modules.

\textbf{M1} translates this foundation into specific embodied AI requirements through world-model-based robotic systems \cite{Zhu_2025_ICCV} and world model-based policy optimization for vision-language-action models \cite{wmpo2026iclr}, defining exactly what infrastructure must protect and coordinate in open-world environments \cite{monwilliams2025embodied}.

\textbf{M2} assesses the quantum threat timeline against these requirements using empirical device data \cite{fedorov2018quantum,nobel2025quantum,turing2026quantum,google2025otoc}, determining the specific security properties and crypto-agility mechanisms that must be maintained in post-quantum migration. 

\textbf{M3} responds with concrete scalable architectures via BrokerChain cross-shard protocols \cite{huang2022brokerchain,huang2025brokerchain,10.14778/3587136.3587143}, demonstrating how to achieve high-throughput coordination while preserving the security guarantees identified in M2. 

\textbf{M4} layers trustworthy data economy mechanisms and interoperability security atop this scalable foundation, implementing Croissant metadata standards \cite{NEURIPS2024_9547b09b} alongside security analysis of cross-ledger communication \cite{augusto2024sok} and blockchain provenance for robotic datasets \cite{yan2025ethbeacon,chemaya2025uniswap}, ensuring that massive data flows from embodied AI systems remain verifiable, incentive-compatible, and secure across heterogeneous ledgers. 

Finally, \textbf{M5} bridges these technical capabilities with industry deployment through cloud infrastructure and multi-modal datasets (e.g., LET, Kuavo\cite{letdataset,kuavochallenge}), demonstrating how the complete stack functions in production environments. 

The \textbf{Closing Panel} synthesizes these components into a cohesive research roadmap, identifying how quantum security, interoperability, and data governance must co-evolve to support next-generation decentralized intelligent environments.





\bibliographystyle{IEEEtran}
\bibliography{references}

\end{document}